\documentstyle[12pt]{article}
\begin{document}

\overfullrule 0 mm
\language 0
\centerline { \bf{  RADIATION REACTION}}
\centerline {\bf {IN CLASSICAL ELECTRODYNAMICS:}}
\centerline { \bf{THE CASE OF ROTATING CHARGED SPHERE}}
\vskip 0.3 cm \centerline {\bf{ Alexander A.  Vlasov}}
\vskip 0.2 cm
\centerline {{  High Energy and Quantum
Theory}} \centerline {{  Department of Physics}} \centerline {{
Moscow State University}} \centerline {{  Moscow, 119899}}
\centerline {{  Russia}}
\vskip 0.2 cm
 03.50.De

{\it In classical electrodynamics for rotating with variable angular
velocity charged rigid sphere  are found: the exact values of
electromagnetic fields, the flux of radiating energy and the exact
integral equation of rotation including the terms of self-rotation.
The analysis of this eq. shows that on one hand, there is no
"runaway" solutions, peculiar to the Lorentz-Dirac eq. of radiating
point particle, on the other hand, there appear new problems, among
them is the nonexistence of solution for some external moments of
force.}

\vskip 0.2 cm
{\bf {1.}}
\vskip 0.2 cm
 In 1998 there will be  60 years since the famous Dirac's  work
on relativistic radiation reaction force in classical electrodynamics
 had appeared. But the problems of relativistic radiation force are
still discussed in the literature ( see [1-10]). Among these problems
are:

(i) can one consider the Lorentz-Dirac equation of motion for
radiating charged point-like particle as the limit  $R\to 0$ of
usual equation of motion for extended charged body   ( $R$ -is the
size of the body)?

(- to  our opinion [8] - the answer is negative: there is no
analyticity near the point $R=0$, thus one cannot mathematically
strictly derive the Lorentz-Dirac equation)

(ii) while radiating, the system must lose its
energy, so has the radiation reaction force the character of damping
force or it can behavior in some cases  as antidamping force? And if
there is antidamping, do there exist runaway solutions of eq. of
motion?

( some  examples of antidamping one can find
in [2,9,10])

To make a careful study of the problem of radiation
reaction one must search for exact solutions of Maxwell equations for
radiating charged systems. Here we consider one such exactly solvable
case - the case of a
charged sphere of radius $R$, rotating with variable angular velocity
$\vec \Omega=\vec \Omega(t)$.
 \vskip 0.2 cm {\bf {2.}} \vskip 0.2 cm

Let the densities of charge and current be:

$$\rho={Q \over 4\pi R^2}\delta(r-R)$$
$$\vec j =[\vec \Omega,\vec r] \rho  \eqno(1)$$
with  $\vec \Omega = \Omega(t) \vec e_{z}$ - i.e. the rotation is the
around the $z$-axis.

(In other words we fixed rigidly at initial moment of time the
surface density of nonrotating rigid sphere to be constant and
after this began the sphere to rotate.)

In Lorentz gauge the eq.  for
electromagnetic potentials $\phi$ and $\vec A$ are (latin indexes
are $0,1,2,3$, greek -$1,2,3$, metric has the diagonal form
$g_{ij}=diag(1,-1,-1,-1)$):  $$ \partial_{p}\partial^{p} \varphi =
4\pi \rho$$ $$ \partial_{p}\partial^{p} \vec A = 4\pi \vec j /c
\eqno(2)$$ Due to the form of current (1), the solution of (2) for
$\vec A$ we can write as $$\vec A =[\vec e_{z}, \vec e_{r}]B(t,r)
\eqno(3)$$ where the function $B$ obeys the eq.:  $${\partial^2 B
\over( c \partial t)^2}-{\partial^2 B \over( \partial r)^2}
-{2\partial B \over r \partial r} + {2 B \over r^2 } = {Q \Omega(t)
 \over cR}\delta(r-R) \eqno(4)$$ With the help of Fourier
transformations:  $$B(t,r)= \int dw \exp{(-iwt)} B^{*}(w,r)$$
$$\Omega(t)= \int dw \exp{(-iwt)} \Omega ^{*}(w)$$ equation (4) for
$B^{*}(w,r)$ takes the form $${\partial^2 \over( \partial
\xi)^2}B^{*}+ {2\partial  \over \xi \partial \xi}B^{*} +(w^2- {2
\over \xi^2 })B^{*} = f \eqno(5)$$ here $\xi = r/c,\ \  \xi_{0} =
R/c$ and $f=f(w,\xi)={Q \over c\xi_{0}}\Omega
^{*}(w)\delta(\xi-\xi_{0}) $

The retarded solution    of (5) is constructed with the
help of Bessel and Hankel functions of semiwhole indexes:
$$B^{*}={\pi \over 2i}\left[ { J_{3/2}(w \xi)\over \xi ^{1/2} }\int
\limits_{\xi}^{\infty} dx x^{3/2} H^{(1)}_{3/2}(wx) +
{H^{(1)}_{3/2}(w \xi)\over \xi ^{1/2} }\int
\limits_{0}^{\xi} dx x^{3/2} J_{3/2}(wx) \right]f(w,x) \eqno(6)$$

The substitution of the inverse Fourier transformation
for $\Omega^{*}$ and  the use  of the
formula $$\int dw \exp {(iwA)}J_{3/2}(wB)H^{(1)}_{3/2}(wC) =$$
$$i{A^2 -B^2-C^2
\over 2 (BC)^{3/2} } \left( \epsilon (A+B+C)-\epsilon (A-B+C) \right)
\eqno(7)$$
with the step-function
$ \epsilon (x) =\left\{ {1, x>0 \atop -1, x<0} \right\}$
(one can derive (7)  using the
standard representation of Bessel and Hankel functions of
semiwhole indexes through the complex exponents)

gives at last for $B(t,r)$ the following result

$$B(t,r)=\left(- {Qc^2 \over 4R r^2} \right) \int
\limits_{t-|r+R|/c}^{t-|r-R|/c} dt' \Omega(t') \left[(t'-t)^2-
(r^2+R^2)/c^2 \right] \eqno(8)$$

The solution for $\varphi$ (1,2) is obvious:
$$\varphi =\left\{ { Q/r, \ \ r>R \atop Q/R , \ \ r<R } \right\}
\eqno(9)$$

Thus we have the exact solutions for the electromagnetic fields $\vec
E= -\nabla \varphi -{\partial \vec A \over c\partial t}$ and $\vec H =
rot \ \vec A$:  $$ \vec E = \vec e_{r}E_{r} + \vec
e_{\phi} E_{\phi},\ \ \ \vec H = \vec e_{r}H_{r} + \vec e_{\theta}
H_{\theta}$$ $$E_{r} = \left\{ {Q/r^2,\ \  r>R \atop 0 ,\ \  r<R }
\right\}, \ \ \ E_{\phi}=-\sin {\theta} {\partial B(t,r) \over c
\partial t}$$ $$H_{r}={2 \cos {\theta} B(t,r) \over r},\ \ \
H_{\theta} =\left(- {\sin {\theta} \over r} \right){\partial (r
B(t,r)) \over \partial r} \eqno(10)$$
\vskip 0.2 cm {\bf {3.}} \vskip
0.2 cm The integration of the energy-momentum balance equation
$$\partial_{j} T^{ij}=0 \eqno(11)$$ over the space volume ( here
$T^{ij}$ - the total (matter+ field) energy momentum tensor) gives
the standard expression for the flux $I$ of radiating energy through
the sphere of radius $r$:
 $$I =\int r^2 d \Xi {c \over 4\pi}(\vec
e_{r} [\vec E, \vec H]) \eqno(12)$$
 here $d\Xi=\sin {\theta} d \theta
d\phi$ -the element of the space angle.  The substitution of
$\vec E, \vec H$ from (10) into eq.(11) yields $$I=-\int r^2
\sin^{3} {\theta}d\theta d\phi {1 \over 4\pi r} {\partial B \over
\partial t}{\partial (rB) \over \partial r} \eqno (13)$$
 In the wave
zone ($r \to \infty$) one can rewrite function $B$ (8) as
$$B\approx
{QR \over 2cr} \Omega(t-r/c)\eqno(14)$$
Consequently, with (14), the energy flux
(13) is expressed as
$$I= {Q^2R^2 \over 6c^2} \left( {\partial \Omega
\over \partial t} \right)^2 $$
 \vskip 0.2 cm {\bf {4.}}
\vskip 0.2 cm
The multiplication of the energy-momentum balance (11) with
$i=\beta$ on $\epsilon_{\alpha \mu \beta}  x^{\mu}$   and
the integration over space volume  $V$ of the rotating sphere yields
the equation of rotation in the form:  $${d \vec N \over dt}= \vec M
\eqno(15)$$
 Here $\vec M$ - is the moment of external forces $\vec M_{ext}$ and
 of electromagnetic field $\vec M_{em}$:  $$\vec M_{em} ={R^3 \over
 4\pi}\int d\Xi\left( [\vec e_{r},\vec E](\vec e_{r},\vec E)+[\vec
e_{r},\vec H](\vec e_{r},\vec H) \right) \eqno(16)$$
and $\vec N$ - is the angular momentum of the rotating rigid sphere
$\vec N_{mech}$ with radius $R$ and total mass $M$ and of the
internal electromagnetic field $\vec N_{em}$ :
  $$\vec N_{mech} = \int dV[\vec r, \vec T_{mech}]$$
$$\vec T_{mech} =  \left( {M \over 4\pi r^2} \right) { [\vec \Omega,
\vec r]\cdot \delta (r-R) \over \sqrt{1- \left(\Omega r \sin
{\theta}/c\right)^2} } \eqno(17) $$ $$\vec N_{em} = \int dV[\vec r,
\vec T_{em}],\ \ \vec T_{em} = {[\vec E, \vec H] \over 4\pi}
\eqno(18)$$ The integrand in (18) for $r \leq R$ (10) is proportional
 to $\vec e_{\phi}E_{\phi}H_{r}$, thus the integration
over $\phi$ yields for it zero result - internal electromagnetic
field gives zero contribution to the equation of rotation.

( Integrating (15) over the infinite sphere ($r\to \infty$) and
taking into account (14), we get the rate of radiation of the total
angular momentum of the system rigid sphere+field: $${d\vec N\over
dt}=- {Q^2R \over 3c^2} {\partial \vec \Omega
\over \partial t} $$ )

 Electromagnetic field (10)for $r=R$  with function $B$ (8), which
we can rewrite as $$B(t,r=R)= \left(- {Qc^2 \over 4R^3} \right) \int
\limits_{t-R/c}^{t} dt' \Omega(t') \left[(t'-t)^2-
{2R^2\over c^2} \right]= $$
$$\left(- {Qc^2 \over 4R^3}  \right)
\int\limits_{0}^{2R/c} dx  \Omega(t-x) (x^2
- {2R^2 \over c^2}) $$
 yields for $\vec M_{em}$ the  result
$$\vec M_{em}= \vec e_{z} \left( {Q^2c \over 6R^2} \right)
\int\limits_{0}^{2R/c} dx {\partial \Omega(t-x) \over \partial t}(x^2
- {2R^2 \over c^2}) \eqno (19)$$
Consequently, after integration of (17), the equation of rotation
(15) takes the form:  $${d \over dt}\left[ {MRc\over \Psi(t)} \left(
{\Psi^2(t)+1 \over 2\Psi(t)}\ln { 1+\Psi(t) \over 1-\Psi(t)} -1
\right) \right]=$$ $${d \over dt} \left[\left( {Q^2c \over 6R^2}
\right) \int\limits_{0}^{2\eta} dx  \Psi(t-x) (x^2 - 2 \eta^2)
\right] + M_{ext} \eqno(20)$$ where $\Psi(t)=\Omega(t) R/c$ and
$\eta=R/c$.

This is nonlinear integro-differential equation (intergal equation
of self-interaction with retardation).  Linearization of eq. (20)
for $\Psi \ll 1$  yields:  $${d \over
dt}\left[ {2MR^2 \over 3} \Omega(t) \right]=$$ $${d \over dt} \left[
 \left( {Q^2c \over 6R^2} \right)\int\limits_{0}^{2\eta} dx
 \Omega(t-x) (x^2 - 2 \eta^2) \right] + M_{ext} \eqno(21)$$ First
terms in R.H.S. of (20-21) describe self-rotation of the sphere.

Thus we see that the self consistent treatment of radiation problem
 implies the non-point-like description of radiating system which
 leads inevitably to intergal equation of retardation (this fact was
 especially stressed in [11]).
\vskip 0.2 cm {\bf {5.}}
\vskip 0.2 cm
If $M_{ext}=0$ then eq.(20-21) are the eq. of self-rotation. Then in
linear approximation eq. (21), rewritten as
$${d \over dt}\left[ k \Omega(t) \right]={d
\over dt} \left[  \int\limits_{0}^{2\eta} dx \Omega(t-x) (x^2 - 2 
\eta^2)
\right] \eqno(22)$$
here $k={4MR^4\over Q^2c},$
has the solution
$${d \over dt}\left[  \Omega(t) \right] =a \exp{bt} \eqno(24)$$
with $a,\ b$- constants, $a$ - arbitrary and $b$ - is the negative
 ($b<0$) solution of the algebraic eq.
$${2MRc^2\over Q^2}={(\nu+1)^2 \over \nu^3}\left( {1-\nu \over 1+\nu}
- \exp{(-2\nu)}\right),\ \ \nu={bR\over c}<-1 $$ This solution
describes the damping of the self-rotation (so there is no runaway
 solutions).

 If $M_{ext}\not= 0$ then the complete solution of (21) is the sum of
 the (24) and of the partial solution of (21). Consequently, to fix
 arbitrary constant $a$, one must use some additional condition,
which physical motivation, generally speaking, is not clear. Partial
solution of (21) can be found by the Laplace transformation of (21),
but as the theory of integral equations tells, not for every
$M_{ext}$ this partial solution exists! Thus we face new problem of
radiation reaction.
\vskip 0.2 cm {\bf {6.}}
\vskip 0.2 cm
Following Lorentz-Dirac approach one can try to extract from eq. (21)
the equation of rotation of point particle with radiation reaction.
For this purpose let us introduce the angular momentum of point
particle $N$ as $N=N(t)= 2MR^2\Omega(t)/3$ and expand (12) in powers
of $R,\ \ R\to 0$. So  if eq.(21) takes the form
${d N(t)\over dt} =$ (infinite with $R\to 0$ term, proportional to
${d N(t)\over dt}$) + (finite with $R\to 0$ term, proportional to
${d^2 N(t)\over dt^2}$), then the first, infinite term leads to
usual regularization, and the second, finite term is what one must
take as radiation reaction.

With $N$ eq. (21) is:
$${d N(t)\over dt}=\lambda {d
\over dt} \left[  \int\limits_{0}^{2\eta} dx N(t-x) (x^2 - 2
\eta^2) \right] + M_{ext} \eqno(25)$$
here $\lambda ={Q^2c\over 4MR^4}$.

Expanding R.H.S. of (25) for $R\to0$ (or $\eta \to 0$), we get:
$$\lambda{d
\over dt} \left[  \int\limits_{0}^{2\eta} dx N(t-x) (x^2 - 2
\eta^2) \right] =$$
$$\lambda{d
\over dt} \left[  \int\limits_{0}^{2\eta} dx \left[N(t)-x{d N(t)\over
dt}+x^2/2{d^2 N(t)\over
dt^2}\right] (x^2 - 2 \eta^2) \right] =$$
$${d N(t)\over dt}\left(-4\eta^3 \lambda/3\right)+{d^3 N(t)\over
dt^3}\left(8\eta^5\lambda/15\right)$$
From this expansion one can see that there is the infinite term,
leading to regularization, but there are no finite with $R\to 0$
terms, so there is no radiation reaction, following Lorentz-Dirac
approach.

Here we once more meet with the incompleteness of Lorentz-Dirac
scheme (see also [8]).

\vskip 0.1 cm {\bf {7.}}
\vskip 0.1 cm

We conclude our treatment with the following remarks:

(i) self-consistent consideration of the problem of radiation
reaction needs the non-point-like description of radiating system;

(ii) the non-point-like description does not coincide with
Lorentz-Dirac scheme for infinitesimally small sizes of the body;

(iii) the non-point-like description of radiation reaction
inevitably leads to integral equation of self-interaction;

(iv) integral equation of self-interaction possesses some peculiar
features, thus solving some old problems of radiation reaction we can
face new ones.

  \vskip 0.2 cm \centerline
 {\bf{REFERENCES}}

  \begin{enumerate}
\item
A.D.Yaghjian, {\it Dynamics of a Charged Sphere}, Lecture notes in
Physics, Springer, Berlin, 1992.
F.Rohrlich, Am.J.Physics, 65(11), 1051 (1997).
 \item S.Parrott, Found.Phys.,23, 1093 (1993).
E.Comay, Found.Phys.,23, 1123 (1993) \item J.Huschilt, W.E.Baylis,
Phys.Rev., D17, 985 (1978).  E.Comay, J.Phys.A, 29, 2111 (1996).
\item J.M.Aguirregabiria, J.Phys.A, 30, 2391 (1997).
J.M.Aguirregabiria, A.Hernandez, M.Rivas, J.Phys.A, 30, L651 (1997).
\item
R.Rivera, D.Villarroel, J.Math.Phys., 38(11), 5630, (1997).
\item W.Troost et al.,  preprint hep-th/9602066.
\item Alexander A.Vlasov, preprint hep-th/9707006.
\item Alexander A.Vlasov, preprint physics/9711024.
\item  E.Glass, J.Huschilt and G.Szamosi, Am.J.Phys.,
52, 445 (1984).
\item Alexander A.Vlasov, Theoretical and Mathematical Physics, 109,
1608 (1996)
\item Anatolii A.Vlasov, {\it Lectures on Microscopic
 Electrodynamics}, Moscow State University, Phys. Dep., 1955
 (unpublished).
  \end{enumerate}

 \end{document}